\documentclass[reprint,aps,pra]{revtex4-1}

\usepackage{graphicx,siunitx,amsmath,mathtools}

\DeclareGraphicsExtensions{.pdf,.png,.jpg}

\begin{document}

\title{Interference with a Quantum Dot Single-Photon Source\\
and a Laser at Telecom Wavelength}

\author{M. Felle}
\affiliation{Toshiba Research Europe Limited, Cambridge Research Laboratory, 208 Cambridge Science Park, Milton Road, Cambridge, CB4 0GZ, United Kingdom}%
\affiliation{Centre for Advanced Photonics and Electronics, University of Cambridge, J.J. Thomson Avenue, Cambridge, CB3 0FA, United Kingdom}

\author{J. Huwer}
\email{jan.huwer@crl.toshiba.co.uk}
\affiliation{Toshiba Research Europe Limited, Cambridge Research Laboratory, 208 Cambridge Science Park, Milton Road, Cambridge, CB4 0GZ, United Kingdom}%

\author{R. M. Stevenson}
\affiliation{Toshiba Research Europe Limited, Cambridge Research Laboratory, 208 Cambridge Science Park, Milton Road, Cambridge, CB4 0GZ, United Kingdom}%

\author{J. Skiba-Szymanska}
\affiliation{Toshiba Research Europe Limited, Cambridge Research Laboratory, 208 Cambridge Science Park, Milton Road, Cambridge, CB4 0GZ, United Kingdom}%

\author{ M. B. Ward}
\affiliation{Toshiba Research Europe Limited, Cambridge Research Laboratory, 208 Cambridge Science Park, Milton Road, Cambridge, CB4 0GZ, United Kingdom}%

\author{I. Farrer}
\affiliation{Cavendish Laboratory, University of Cambridge, J.J. Thomson Avenue, Cambridge, CB3 0HE, United Kingdom}%

\author{R. V. Penty}
\affiliation{Centre for Advanced Photonics and Electronics, University of Cambridge, J.J. Thomson Avenue, Cambridge, CB3 0FA, United Kingdom}

\author{ D. A. Ritchie}
\affiliation{Cavendish Laboratory, University of Cambridge, J.J. Thomson Avenue, Cambridge, CB3 0HE, United Kingdom}%

\author{A. J. Shields}
\affiliation{Toshiba Research Europe Limited, Cambridge Research Laboratory, 208 Cambridge Science Park, Milton Road, Cambridge, CB4 0GZ, United Kingdom}%

\date{\today}

\begin{abstract}

The interference of photons emitted by dissimilar sources is an essential requirement for a wide range of photonic quantum information applications.
Many of these applications are in quantum communications and need to operate at standard telecommunication wavelengths to minimize the impact of photon losses and be compatible with existing infrastructure.
Here we demonstrate for the first time the quantum interference of telecom-wavelength photons from an InAs/GaAs quantum dot single-photon source and a laser; an important step towards such applications.
The results are in good agreement with a theoretical model, indicating a high degree of indistinguishability for the interfering photons.

\end{abstract}

\maketitle

Single-photon sources are essential components for many photonic quantum information technologies, ranging from linear-optics quantum computation \cite{Knill2001N} to teleportation of quantum bits \cite{Bennett1993PRL} and large scale quantum networks. \cite{Kimble2008N}
The interference of two independently generated photon states on a beam splitter is an important physical mechanism required for the realization of most of these schemes.

Apart from the common implementation with two identical single photons, \cite{Hong1987PRL} great potential lies in the interference of states with completely different statistical properties, such as a single-photon Fock state and a weak coherent state.
It has been demonstrated that this provides a versatile tool to fully characterize the spectral \cite{Wasilewski2007PRL} and temporal \cite{Qin2015NLSA} density matrix of unknown single-photon states.
Further applications are quantum amplifier schemes,\cite{Josse2006PRL} suitable for quantum-noise limited amplification of coherent states.

Of even greater immediate importance are applications related to quantum communication and quantum key distribution \cite{Ekert1991PRL, Gisin2002RMP} (QKD), the most developed technology based on photonic quantum bits.
The most widely implemented scheme for QKD makes use of weak coherent laser pulses.\cite{Hwang2003PRL}
Interference of these states with single photons from an entangled pair to perform a Bell-state measurement thereby enables quantum teleportation.
This opens up the route to develop a so-called quantum relay \cite{Jacobs2002PRA} or all-photonic quantum repeater, \cite{Azuma2015NC} indispensable to reduce noise and extend the transmission distances in future networks that are strongly limited by photon losses in optical fiber.
More generally, recent theoretical studies \cite{Andersen2013PRL} have shown that, for limited experimental resources, a hybrid approach for the teleportation of continuous variable systems by using discrete single-photon entangled states is expected to have significant advantages over its continuous-variable counterpart.
The interference between dissimilar photon sources is thereby of great interest both for fundamental science and a large number of technological applications.

Most experiments demonstrated so far have been performed with heralded single photon sources based on non-linear optical processes. \cite{Rarity2005JOB, Wasilewski2007PRL, Qin2015NLSA}
These sources obey Poissonian statistics, intrinsically deteriorating the single-photon character and making them non-desirable for certain applications.
More recently, quantum dots (QDs) based on III-V semiconductor compounds have proven to be one of the most promising sub-Poissonian sources, generating deterministic single photons as well as entangled photon pairs. \cite{Benson2000PRL, Michler2000S, Stevenson2006N, Muller2014NP}
In the past, these systems have been successfully used to demonstrate interference of single photons with emission from a laser \cite{Bennett2009NP} and subsequently the teleportation of laser-generated qubits. \cite{Stevenson2013NC}
In both cases, the operating wavelength was below 1$\SI{}{\micro\meter}$, making these sources unsuitable for quantum communication applications due to high photon absorption in optical fiber.

Operation of single photon sources in the standard O-band ($\sim$1310\,nm) and C-band ($\sim$1550\,nm) telecommunication windows provides much lower losses and compatibility with existing optical-fiber-based communication infrastructures. \cite{Townsend1997N,Patel2012PRX}
Over the past decade, progress has been made in increasing the emission wavelength of QD devices from the near infrared to standard telecom wavelengths, \cite{Alloing2005APL, Ward2005APL, Benyoucef2013APL} including the direct generation of single entangled photon pairs in the telecom O-band. \cite{Ward2014NC}
To date, there is no reported interference measurement with this new class of quantum light emitter.
In this Letter, we demonstrate for the first time the quantum interference between single photons generated at telecom wavelength from a QD with photons emitted from a laser.

\begin{figure}[t]%
\includegraphics[clip,trim=0.25cm 0 0.45cm 0.25cm,width=\columnwidth]{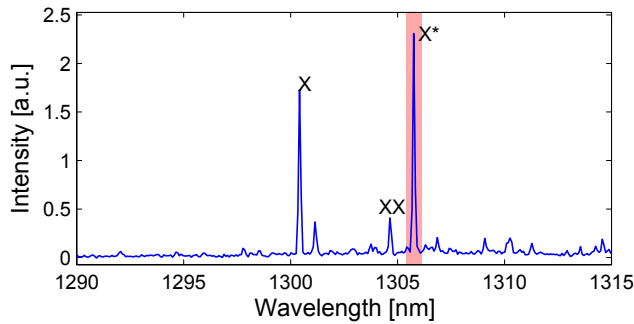}%
\caption{Photoluminescence spectrum of QD when exciting with 785\,nm laser light at 10\,K for a bias voltage of +1\,V. The shaded area denotes the part of the spectrum filtered for our experiment, centered around a bright and unpolarized charged exciton line (X*).}
\label{fig:spectrum}%
\end{figure}

The sample contains InAs/GaAs QDs in a quantum well, at the center of the intrinsic region of a \textit{p-i-n} diode, surrounded by a weak planar distributed Bragg reflector cavity made from stacks of AlGaAs/GaAs, grown on a GaAs substrate.  \cite{Ward2014NC}
Applying a bias voltage across the diode tunes the emission properties of the device under optical excitation. 

The device was operated at 10\,K, and optically excited with continuous wave laser light at 785\,nm. 
Unless otherwise stated in this paper, the applied bias voltage was set to +1\,V.
Emitted photons from a single QD were collected with an aspheric lens (NA = 0.55) and coupled to a single-mode fiber. Photoluminescence spectra were measured with a grating spectrometer equipped with an InGaAs detector array.

Figure \ref{fig:spectrum} shows the spectrum of the single QD used in this work. The labelled lines were identified by a combination of power-dependency measurements with the exciting laser and fine structure splitting (FSS) measurements based on the polarization properties of emitted photons.
We deduce that the FSS is $88\pm3$\,$\SI{}{\micro\electronvolt}$ for the X and XX line.
The charged exciton line exhibits no measurable splitting and emits unpolarized photons.
The QD was selected for its bright emission, with separated spectral lines, and good coherence properties.

\begin{figure}[t]%
\includegraphics[width=\columnwidth]{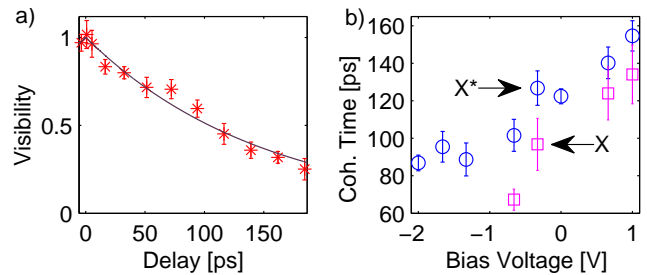}%
\caption{a) Measurement of single photon coherence time for the charged exciton line at 1306\,nm from Fig. \ref{fig:spectrum}. The solid line is an exponential fit, corresponding to a coherence time of $150\pm9$\,ps. b) Measured coherence time of the charged (X*) and neutral (X) exciton from the QD as a function of the applied bias voltage to the device.}%
\label{fig:cohtime}%
\end{figure}

Two-photon interference experiments require long photon coherence times to achieve highest visibilities in experimental setups with limited timing resolution.
We used a fiber-based Mach-Zehnder interferometer to characterize the coherence properties of emitted photons from QDs in the sample. 
Figure \ref{fig:cohtime} a) shows the measured single-photon interference visibility as a function of the delay in the interferometer, for photons emitted on the charged exciton line at 1306\,nm. 
We derive a coherence time of $\tau_c=150\pm9$\,ps from the exponential fit, for a bias voltage of $+1$\,V applied to the device. 

Embedding a QD in a vertical \textit{p-i-n} diode \cite{Benson2000PRL} provides a useful tool to control the emission properties by application of an electric field across the dot layer.
This control ranges from tunability of the emission wavelength and FSS via the quantum confined Stark effect, \cite{Bennett2010NP,Marcet2010APL} to the direct electrical excitation of QDs.  \cite{Yuan2002S}
In addition, there is a significant influence on the coherence time of the dot emission. 
Figure \ref{fig:cohtime} b) shows a measurement of coherence times of the two brightest emission lines from the QD at different bias voltages. 
We observe an almost two-fold increase in the coherence time when changing the bias voltage from $-2$\,V to $+1$\,V, approaching the flat-band operation, which is expected around $+2.2$\,V.  \cite{Ward2014NC}
This effect is most probably caused by an effective relaxation of the fluctuating charge environment surrounding the QD for an increasing flow of carriers injected into the device.
The application of bias voltages greater than $+1$\,V resulted in a reduced brightness of the emission and no further extension of the coherence time. 

\begin{figure}[t]%
\includegraphics[width=\columnwidth]{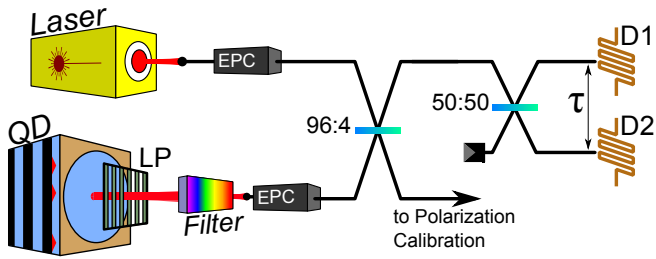}%
\caption{Diagram of the experimental setup used to measure two-photon interference. LP: linear polarizer, EPC: electronic polarization controller, D1/D2: superconducting nanowire single photon detectors.}%
\label{fig:setup}%
\end{figure}

We then used single photons emitted from the charged exciton line at 1306\,nm to measure two-photon interference with photons generated by a laser. 
We implemented an all-fiber interference setup with an unbalanced beam splitter to keep the losses of QD photons low.
Figure \ref{fig:setup} gives an overview of the main components. 
The QD emission is sent through a linear polarizer and a narrow-band spectral filter, providing photons with well-defined polarization and frequency. 
The photons are overlapped with a strongly attenuated laser beam in a 96:4 single-mode fiber beam splitter (BS) made from standard telecom fiber (SMF-28). 
Autocorrelation measurements of the output mode carrying 96\,\% of incident QD photons and 4\,\% of incident laser photons are then performed with a Hanbury Brown and Twiss setup \cite{HBT1956N} comprising a balanced 50:50 fiber BS, two superconducting nanowire single photon detectors (Single Quantum) and time-correlated single photon counting electronics.
The ratio between the rates of detected laser photons and detected QD photons is controlled by adjusting the laser intensity and is typically set to a value around 0.5.
The second output mode of the unbalanced fiber BS is used to calibrate the polarizations of the two interfering modes, which are controlled by electronic polarization controllers (EPCs). 
The polarization of QD photons is switched between the parallel and crossed case with respect to the laser polarization, thereby enabling the measurement of interfering and non-interfering photons. 
Laser photons are generated from a commercial diode laser, tunable across the telecom O-band with a precision of $<$1 $\SI{}{\micro\electronvolt}$. 

To observe perfect Hong-Ou-Mandel interference, \cite{Hong1987PRL} the photons must be indistinguishable in both their polarization and frequency. 
We tuned the laser wavelength to the QD emission by overlapping both spectral lines on the spectrometer.
Gaussian fits were used to overcome the spectrometer’s resolution of  $\sim\SI{60}{\micro\electronvolt}$, achieving a match of both energies with a precision of $\pm\SI{2}{\micro\electronvolt}$. 
Assuming that the QD photons are Fourier-limited, the measured coherence time of \SI{150}{\pico\second} corresponds to a spectral bandwidth of $\sim\SI{4 }{\micro\electronvolt}$, sufficiently large to provide good indistinguishability of the two photon sources.

\begin{figure}[t]%
\includegraphics[trim=0 0.2cm 0 0.8cm,clip,width=\columnwidth]{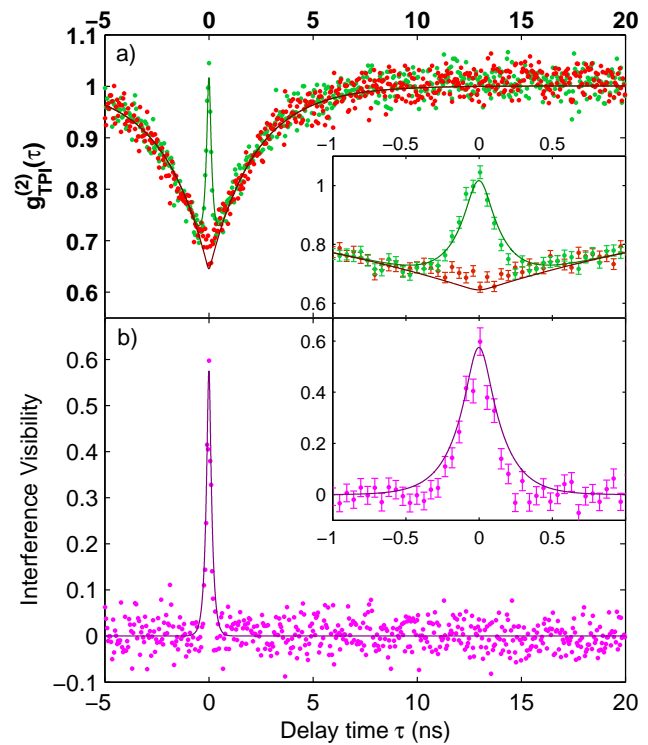}%
\caption{a) Correlation measurement of two-photon interference for distinguishable (red) and indistinguishable (green) photons for a bin size of 48\,ps.
For the sake of clarity Poissonian errors are only displayed in the inset. The solid lines show the theoretical model described in Equation \ref{eq:TPI}, convoluted with a Gaussian detector response. b) Interference visibility calculated from the coincidence measurements. Error bars in the inset are the propagated errors for each time bin. The purple line shows the calculated visibility from the model displayed in a).}%
\label{fig:Result}%
\end{figure}

From a $g^{(2)}$ measurement of the QD emission, we derive a $g_{\mathrm{QD}}^{(2)}(0)$ value of $21\pm4$\,\%, proving the sub-Poissonian character of the source.
This value is limited by detector dark counts, timing jitter of the electronics and background collected from neighboring QDs in the sample.

Previous interference experiments using dissimilar photon sources \cite{Bennett2009NP, Stevenson2013NC} have shown that the measurements are well understood when using the wavepacket description developed in \citet{Legero2003APB}.  
For the unbalanced interference circuit described in Fig. \ref{fig:setup}, the theoretical second-order correlation function is
\begin{align}
	g^{(2)}_{\mathrm{TPI},\phi}(\tau)=1+\frac{\eta^2 \left(g^{(2)}_{\mathrm{QD}}(\tau)-1\right)+2 \eta \alpha^2\mathrm{e}^{-\left|\tau\right|/\tau_c}\cos^2\phi }{\left( \eta+\alpha^2+\beta\right)^2}
	\label{eq:TPI}
\end{align}
with $\eta$, $\alpha^2$, and $\beta$ being proportional to the separate QD, laser, and background photon intensities measured at detectors D1 and D2. 
The interference visibility is defined as $V(\tau)=[ g^{(2)}_{\parallel}(\tau)-g^{(2)}_{\perp}(\tau) ]/ g^{(2)}_{\perp}(\tau)$, where the subscripts $\parallel$ and $\perp$ refer to co- and cross-polarized measurements of $g^{(2)}_{\mathrm{TPI}}$ ($\phi=0$ and $\pi/2$ in Equation \ref{eq:TPI}). 
Taking the timing jitter and the previously determined single photon properties of the QD emission into account, we expect a maximum visibility of the interference for a laser to QD emission intensity ratio of $\alpha^2/\eta\simeq0.5$.

Figure \ref{fig:Result} a) shows the measured normalized coincidences, without background subtraction, after collecting data for co- and cross-polarized photons. 
For non-interfering photons, the second-order correlation function $g^{(2)}_{\perp}(\tau)$ shows an anti-bunching dip originating from the single-photon source, superposed on the uncorrelated laser-laser and dot-laser photon coincidences. 
For the interfering case, $g^{(2)}_{\parallel}(\tau)$ exhibits a narrow and strong bunching peak at $\tau$=0 due to the coalescence of indistinguishable photons through the same output mode of the unbalanced BS. 
The lines show the theoretical fit from Equation \ref{eq:TPI} convoluted with a Gaussian detector response, accounting for a measured timing jitter of $101.9\pm0.4$\,ps.
The single-photon coherence time $\tau_c$ was determined in an independent measurement, as shown in Fig. \ref{fig:cohtime} a).
We observe very good agreement between the model and the experimental data.
The slight discrepancy for the non-interfering case can be explained by a deviation from perfectly crossed polarizations, caused by the limited extinction ratio of the fiber components, used for calibration.

The interference visibility available directly from the raw data is displayed in Fig. \ref{fig:Result} b), with a maximum measured value of $60\pm6$\,\%.
It has to be emphasized that this class of interference experiment between dissimilar sources is dominated by the Poissonian nature of the laser, limiting the visibility due to random laser-laser coincidences, depending on the laser intensity with respect to the QD emitter.
By using Equation \ref{eq:TPI} and the experimental intensity ratio of $\alpha^2/\eta=0.63\pm0.04$ we calculate a theoretical visibility of 76.2\,\%, assuming no timing jitter or detector dark counts and a perfect single-photon emitter.
Therefore, the measured raw data visibility corresponds to $79\pm8$\,\% of the maximal achievable value for the chosen intensity ratio of both sources.

To conclude, we have for the first time demonstrated the interference of single photons emitted by a semiconductor QD at the center of the telecom O-band with photons generated by a laser. 
The resulting raw-data interference visibility of 60\,\% compares well
with raw-data visibilities achieved in other two-photon interference experiments between identical QD single-photon sources, operating at lower wavelength. \cite{Muller2014NP}
The fact that we observe good agreement with our theoretical model is a strong indication of a high degree of indistinguishability between the two independent photon sources.
Even higher visibilities will be achievable by improving the single-photon properties of the source through the application of resonant \cite{Stufler2006PRB,Muller2014NP} or quasi-resonant \cite{Intallura2007APL} optical excitation schemes, or different growth technologies. \cite{Paul2015APL}
Assuming an overall fidelity of 85\,\% for an entangled photon pair source, as recently demonstrated at telecom wavelengths, \cite{Ward2014NC} the achieved visibility would be sufficient to implement quantum teleportation with fidelities above 80\,\%, guaranteeing security for QKD applications. \cite{Chau2002PRA}
The presented results are an important step towards a large number of photonic quantum information applications at telecom wavelength.
Most importantly, they pave the way for the implementation of a high fidelity quantum relay compatible with available telecommunication infrastructure.

\begin{acknowledgments}
The authors acknowledge partial financial support from the Engineering and Physical Sciences Research Council, and the EPSRC Quantum Technology Hub in Quantum Communications. MF acknowledges support from the EPSRC CDT in Photonic Systems Development.
\end{acknowledgments}

\end{document}